\documentstyle[aps,twocolumn,floats,epsfig]{revtex}

\begin{document}
\title{$d+is$ vs $d+id^{\prime}$ Time Reversal Symmetry Breaking States
in Finite Size Systems}
\author{M.H.S. Amin$^1$, S.N. Rashkeev$^2$, M. Coury$^1$,
A.N. Omelyanchouk$^3$, and A.M. Zagoskin$^{1,4}$}
\address{$^1$D-Wave Systems Inc., 320-1985 W.~Broadway, Vancouver, B.C.,
V6J 4Y3, Canada}
\address{$^2$Dept.~of Physics and Astronomy, Vanderbilt
University, Box 1807 Station B, Nashville, TN 37235, USA}
\address{$^3$B.I.Verkin Institute for Low Temperature Physics and
Engineering, Ukrainian National Academy of Sciences, \\
Lenin Ave.~47, Kharkov 310164, Ukraine}
\address{$^4$Physics and Astronomy
Dept.,~The University of British Columbia, \\
6224 Agricultural Rd., Vancouver, B.C.,~V6T 1Z1, Canada}
\maketitle

\begin{abstract}
We report self-consistent quasiclassical calculations of
spontaneous currents and magnetic moments in finite size
unconventional superconducting systems, namely: (i) in isolated
$d$-wave superconductor islands where, in addition to the
dominant order parameter (with a $d_{x^2-y^2}$ symmetry), a
subdominant order parameter of $s$ or $d_{xy}$ symmetry is added;
(ii) in grain boundary junctions between two arbitrarily oriented
$d$-wave superconductors, and between a $d$-wave and an $s$-wave
superconductor. We show that the profile of the spontaneous
current density and the magnetic field distribution depend on the
time-reversal symmetry breaking properties of the system. For the
$d_{x^2-y^2}+id_{xy}$ state, vortices appear near the edges of
the finite size systems. We associate these vortices with the
chiral nature of the mixed order parameter. The method developed
here is quite general, and can be used for predicting properties
of any finite size superconducting system.
\end{abstract}


\vspace{-1cm}

The possibility of the existence of mixed order parameter
symmetries in unconventional superconductors has been suggested
and recently investigated both theoretically
\cite{sigrist,huck,barash,amin,amin1,matsumoto,Yip,Fogelstrom} and
experimentally
\cite{Covington,mannhart,NV01,carmi,tafuri,walter}. A significant
feature of the mixed symmetry states is that they may produce
spontaneous currents and magnetic moments which can be measured
using appropriate experimental techniques. High $T_c$ cuprates
present an especially interesting class of materials for studying
mixed symmetry states. In these systems, recently developed
technology \cite{tzalenchuk} provides an opportunity to fabricate
different structures with controllable characteristics, such as
$\pi$-junctions \cite{sigrist,pjn} (junctions with equilibrium
phase difference equal to $\pi$), submicron size
$\phi_{0}$-junctions \cite{ilichev} (with an equilibrium phase
difference $\phi_0$ which is different from 0 or $\pi$
\cite{amin1}), $\pi$-SQUIDs \cite{sigrist,schulz} and
$\pi/2$-SQUIDs \cite{ACR}, and superconducting qubit prototypes
\cite{qubit}. Therefore, a quantitative prediction of properties
of such finite size unconventional superconducting systems
becomes extremely important for both experiments and technology.

The mixed symmetry states can be realized near a surface of a
$d$-wave superconductor or at a grain boundary junction between
two different superconductors. In the grain boundary junction
case, the symmetry of the mixed state is dictated by the
symmetries of the order parameters on both sides of the junction
(proximity effect)\cite{sigrist,huck,barash,amin,amin1}. On the
other hand, at a surface, a subdominant order parameter may
appear due to an attractive interaction potential in the
corresponding channel. This may involve a second order phase
transition below the superconducting transition temperature \cite
{amin1}. The mixed symmetry states that have been suggested for
$d$-wave superconductors are $d_{x^2-y^2}+id_{xy}$ and
$d_{x^2-y^2}+is$ \cite{matsumoto} (we denote them $d+id^{\prime}$
and $d+is$, respectively). Which of these states is realized near
the surface of a $d$-wave superconductor is still an open
question. An identification of this state would provide useful
information about microscopic interactions in the superconductor.
The $d+id^{\prime}$ and $d+is$ states are inherently different in
the sense that the Cooper pairs in the former case have intrinsic
magnetic moment, while in the latter, they do not
\cite{volovik,amin1}. Therefore, it is natural to expect
differences in some measurable quantities, such as spontaneous
current and magnetic field.

In this article, we perform self-consistent calculations of
spontaneous current densities and magnetic field distributions
for finite size systems with different mixed symmetries. We show
that these experimentally observable characteristics are very
dependent on the time-reversal symmetry properties of the system.
For the $d+id'$ state, we find that vortices appear, and we
connect their appearance with the chiral nature of the order
parameter in this state.

Most general approaches to calculation of currents in
superconducting systems are based on Gorkov equations for Green's
functions of the superconductor. It is widely accepted that this
``mean field'' approach is valid on a phenomenological level
\cite{book}, independent of further developments in the first
principles' theory of high $T_{c}$ superconductivity. In a
quasiclassical limit, these equations (which are also called
Eilenberger equations, see Ref. \onlinecite{Eilenberger}) can be
solved by integrating along the classical trajectories of the
quasiparticles (with boundary conditions at infinity). In finite
size systems, where the trajectories undergo multiple reflections
from surfaces and interfaces, the problem of definition of the
boundary conditions becomes complicated. We demonstrate how a
modification of the Eilenberger formalism based on the
Schopohl-Maki transformation \cite{Schopohl} can be used for
stable numerical calculations in finite size two-dimensional (2D)
systems.

The Eilenberger equations for quasiclassical Green's functions
can be written as
\begin{equation}
{\bf v}_{F}\cdot \nabla \widehat{g}+[\omega \widehat{\tau }_{3}+
\widehat{\Delta },\widehat{g}]=\widehat{0}, \qquad \widehat{g}^{\
2}=\widehat{1}, \label{eil}
\end{equation}
where $\omega $ is the Matsubara frequency and
\[
\widehat{\tau}_3=\left(
\begin{array}{cc}
1  & 0  \\
0 & -1
\end{array}
\right) ,\quad \widehat{g}=\left(
\begin{array}{cc}
g  & f  \\
f ^{\dagger } & -g
\end{array}
\right) ,\quad \widehat{\Delta }=\left(
\begin{array}{cc}
0 & \Delta \\
\Delta ^{\dagger } & 0
\end{array}
\right) .
\]
The matrix Green's function $\widehat{g}$ and the superconducting
order parameter $\Delta$ are both functions of the Fermi velocity
${\bf v}_{F}$ and the position ${\bf r}$. $\Delta$ is determined by
the self-consistency equation, which in
the two-dimensional (2D) case can be written as
\begin{eqnarray}
\Delta (\theta)=2\pi N(0)T\sum\limits_{\omega
>0} \left< V_{\theta\theta'}f (\theta')
\right>_{\theta'}
\label{DDD}
\end{eqnarray}
where $\theta$ is the angle between ${\bf v}_F$ and the $x$-axis,
$V_{\theta\theta'}$ is the interaction potential, $N(0)$ is the
density of states at the Fermi surface, and
$\left<...\right>_\theta$ represents averaging over $\theta$. We
also have $\Delta^\dagger = \Delta^*$, which holds for any singlet
order parameter symmetry. Generally, it is possible to obtain an
order parameter which is a mixture of several terms with different
symmetries, e.g., $\Delta =\Delta _{x^{2}-y^{2}}+\Delta
_{xy}+\Delta _{s}$, where $\Delta _{x^{2}-y^{2}}=\Delta _{1}\cos
2\theta ,$ $\Delta _{xy}=\Delta _{2}\sin 2\theta $, and $\Delta
_{s}$ are the dominant $d_{x^{2}-y^{2}}$ component, the
subdominant $d_{xy}$, and the $s$ components of the order
parameter, respectively. The corresponding interaction potential,

\begin{equation}
V_{\theta \theta ^{\prime }}=V_{d1}\cos 2\theta \cos 2\theta
^{\prime }+V_{d2}\sin 2\theta \sin 2\theta ^{\prime }+V_{s},
\label{V}
\end{equation}
should be substituted in the self-consistency equation (\ref{DDD})
in order to assure a self-consistency of the
solution~\cite{amin1}. The current density ${\bf j(r)}$ is given
by
\begin{eqnarray}
{\bf j}=-4\pi ieN(0)T\sum\limits_{\omega >0} \left< {\bf v}_{F}g
\right>_{\theta }. \label{j}
\end{eqnarray}

For numerical calculations, it is convenient to parameterize the
quasiclassical Green's functions via \cite{Schopohl}

\begin{equation}
g={\frac{1-ab}{1+ab}}\ ,\quad f={\frac{2a}{1+ab}}\ ,\quad
f^{\dagger }=\frac{2b}{1+ab}.  \label{EqC1}
\end{equation}
Functions $a$ and $b$ satisfy two independent, but nonlinear,
(Riccati) equations:

\begin{eqnarray}
{\bf v}_{F}\cdot \nabla a &=&\Delta -\Delta ^{\ast }a^{2}-2\omega
a, \nonumber \\
-{\bf v}_{F}\cdot \nabla b &=&\Delta ^{\ast }-\Delta b^{2}-2\omega b.
\label{EqC2}
\end{eqnarray}
From these equations it follows that $a(-{\bf v}_{F})=b^{\ast
}({\bf v}_{F})$ and $b(-{\bf v}_{F})=a^{\ast }({\bf v}_{F})$. The
solutions of these equations can be interpreted as {\it
trajectories} of quasiparticles. One should integrate these
equations along all possible trajectories and perform the
summation over the trajectories to calculate the current. To find
$a$ and $b$ along the trajectories, one needs to use boundary
conditions at the ends of the trajectories. In infinite systems,
one usually assumes that {\em all} trajectories go deep into the
bulk of the superconductor, i.e., one can use the bulk solutions

\begin{equation}
a_{\pm }={\frac{\Delta }{\omega \pm \Omega }}\ ,\qquad b_{\pm
}={\frac{\Delta ^{\ast }}{\omega \pm \Omega }} ,  \label{EqC3}
\end{equation}
with $\Omega =\sqrt{\omega ^{2}+|\Delta |^{2}}$, as the boundary
conditions at infinity. In the case of finite size systems where
multiple reflections from surfaces and interfaces are possible,
stability of the numerical procedure for calculating the current
is not obvious because of complications with choosing a proper
boundary condition for a given trajectory. Nevertheless, we show
that a stable numerical procedure can still be developed.

Numerical integration for $a$ ($b$) in Eq.~(\ref{EqC2}) should be
taken in the direction of ${\bf v}_{F}$ (-${\bf v}_{F}$) to
ensure stability. When $\Delta $ is a constant, the solution for
$a$ can be written analytically,

\begin{eqnarray}
a_{f} &=&a_{+}+{\frac{a_{i}-a_{+}}{1+{\frac{\Delta ^{\ast
}}{\Omega }}
(a_{i}-a_{-})e^{\Omega \tau }\sinh \Omega \tau }} \\
&\approx &a_{+}+{\frac{\Omega }{\Delta ^{\ast }}}\left(
{\frac{a_{i}-a_{+}}{a_{i}-a_{-}}}\right) e^{-2\Omega \tau }
\qquad ({\rm if}\ \Omega \tau \gg 1), \nonumber
\end{eqnarray}
where $a_{i}$ and $a_{f}$ are the values of $a$ at the initial
(${\bf r}_{i}$) and final (${\bf r}_{f}$) points of the
trajectory, and $\tau =|{\bf r} _{f}-{\bf r}_{i}|/v_{F}$ is the
migration time between the initial and final points. It is clear
that the solution for $a$ relaxes to the bulk value $a_{+}$ at
the distance $L={\bf v}_{F}/2\Omega $ which is of the order of
the coherence length $\xi _{0}$. In other words, when the
quasiparticle moves away from the initial point at a distance of
a few $\xi _{0}$'s, any information about the initial point
$a_{i}$ is lost. The same argument is valid for the function $b$.

Let us now consider a system with a restricted geometry. After
integrating over a few $\xi _{0}$'s (considering the
reflections), $a_{f}$ becomes almost independent of $a_{i}$. This
solution corresponds to a simple exponential relaxation of the
functions $a$ and $b$ to their local ``steady-state'' values
(which can be different from the bulk values) defined by the
local values of the order parameter. Therefore, to find $\Delta$
at a given point, one does not need the values of $\Delta$ at
distances larger than several $\xi _{0}$ along the trajectory.
Such a ``relaxational'' property of the $a$ and $b$ functions
significantly simplifies the numerical solution of the
self-consistent 2D problem. All the ``multiple-reflection
history'' of a trajectory becomes a moot point, and for practical
calculations, this trajectory can be cut at a distance of a few
$\xi_{0}$'s from a given point. We used this observation in our
self-consistent calculations. For the ``cutting'' distance, we
choose $10\xi _{0}$--$20\xi _{0}$. We set the bulk values of $a$
and $b$ ($a_{+}$ and $b_{+}$) calculated at the initial point of
the ``truncated'' trajectory as the boundary conditions (such a
choice does not affect the final results because the system has
no memory) and integrate along the trajectory until we get to the
point where we calculate the current. We found that the results
are very stable and do not depend on the value of the ``cutting''
distance.

\begin{figure}[t]
\epsfysize 5.5cm \epsfbox[0 50 350 750]{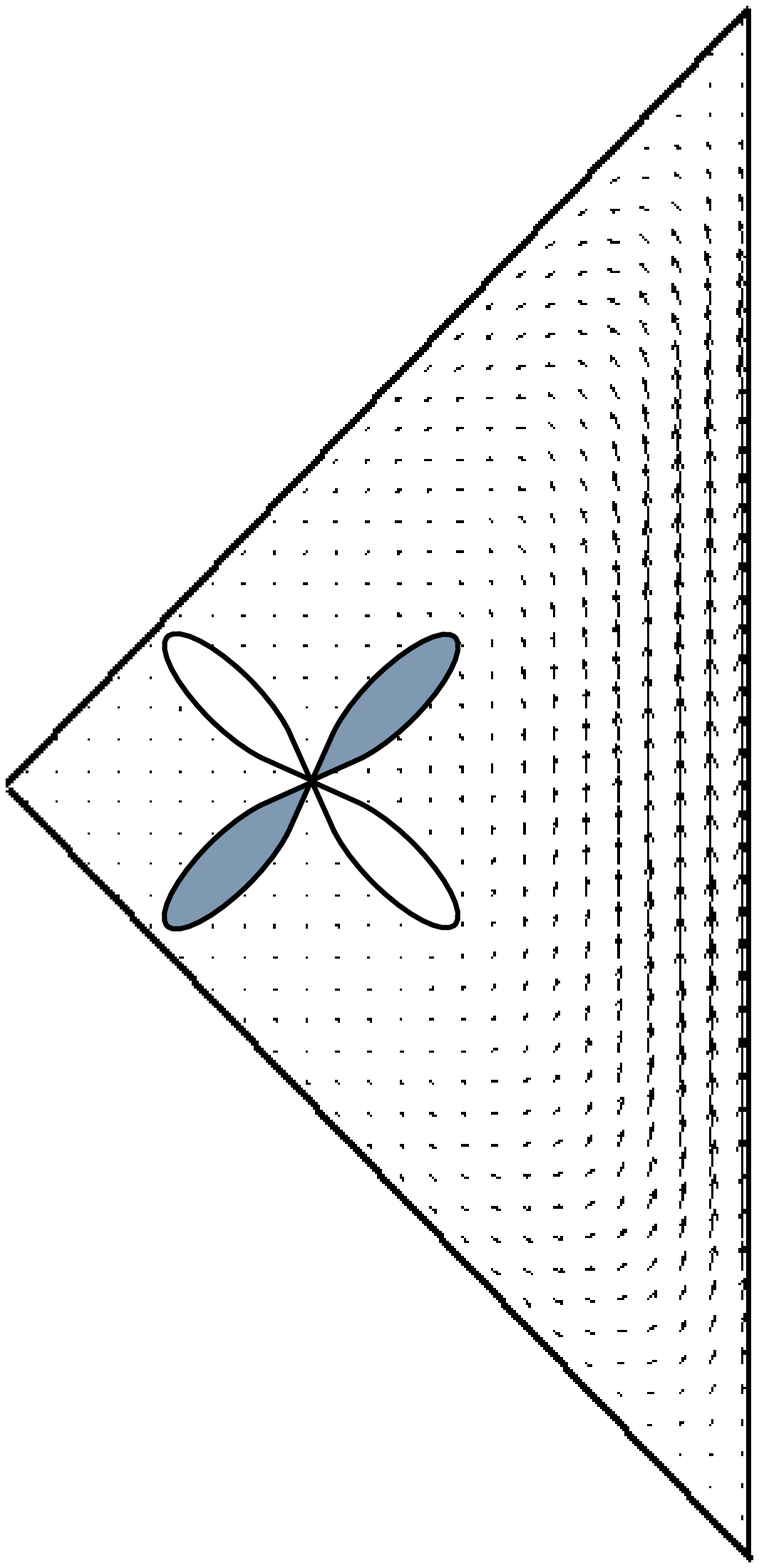} \epsfysize 5cm
\epsfbox[100 220 420 620]{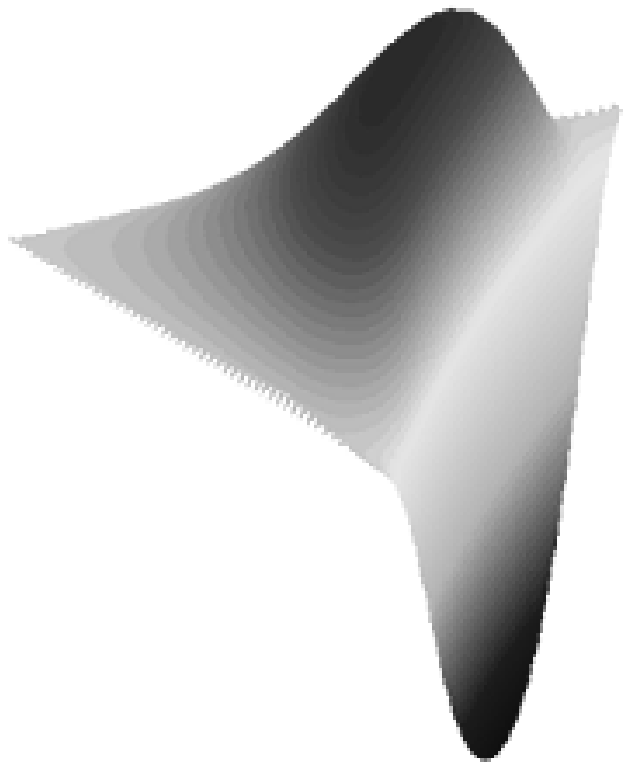} \caption{{\protect\small
Spontaneous current (left) and magnetic field (right) in a finite
size triangular-shaped region of a $d$-wave
superconductor with a subdominant $s$-wave, at the temperature $%
T=0.2T_{cs}=0.1T_c$. The orientation of the dominant
$d_{x^2-y^2}$ order parameter is shown.}} \label{tris}
\end{figure}

\begin{figure}[t]
\epsfysize 5.5cm \epsfbox[0 50 350 750]{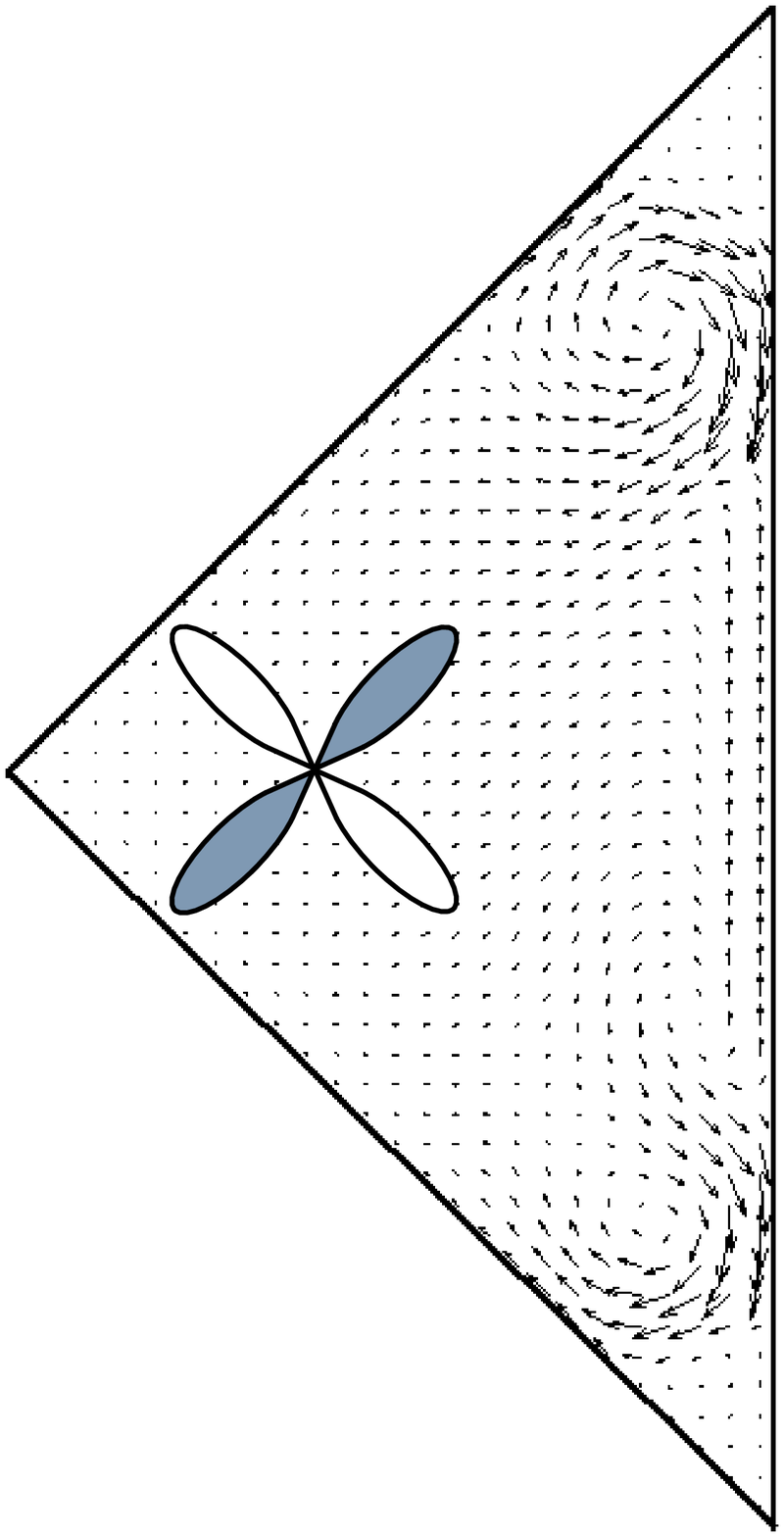} \epsfysize 5cm
\epsfbox[100 220 420 620]{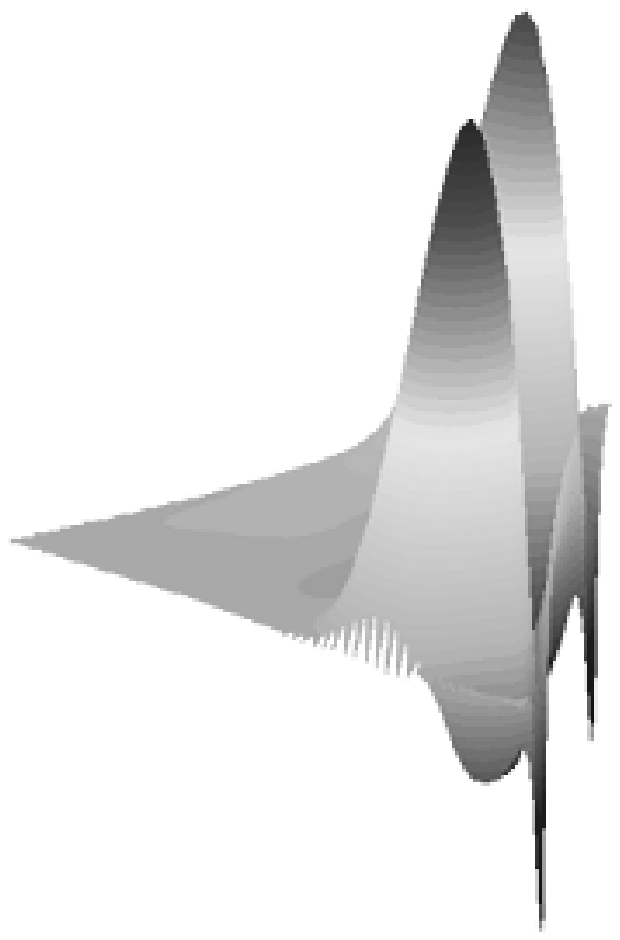} \caption{{\protect\small The
results of self-consistent calculations similar to those of
Fig.~\ref{tris} but with a subdominant $d_{xy}$-wave order
parameter, and at the temperature $T=0.2T_{c2}=0.1T_c$.}}
\label{trid}
\end{figure}

To compare the spontaneous currents and magnetic fields in the
$d+is$ and $d+id^{\prime}$ states, we performed self-consistent
calculations of the order parameter in small size triangular
regions of clean $d$-wave superconductor with different
subdominant order parameters. We assume specular boundaries with
the length of the longer edge of the triangles being $30 \xi_0$.
Figs.~\ref{tris} and \ref{trid} show the spontaneous current and
magnetic field distributions in systems with the two different
symmetries. We imposed the desired subdominant symmetry by
introducing an additional attractive interaction potential in the
corresponding channel [$V_{d2}$ or $V_s$ in Eq. (\ref{V})]. The
subdominant critical temperatures are taken to be $0.5T_c$ in
both cases.  The orientation of the dominant order parameter
makes a $45^\circ$ angle with respect to the longer (right) edge
of the triangle (see the figures). As a result, only this
boundary of the triangle is pair breaking; the quasiparticles
face a different sign of the dominant order parameter after
reflection from the surface \cite{amin1}. The suppression of the
main order parameter at the surface allows the subdominant order
parameter to appear near the surface.

The spontaneous current and magnetic field distributions are
evidently different in Figs.~\ref{tris} and \ref{trid}. In the
first case, the current makes a counterclockwise loop throughout
the region, while in the second case, the direction of rotation
of the current changes throughout the triangle. Especially,
appearance of the two vortices in Fig.~\ref{trid} reflects the
chiral nature of the $d+id^{\prime}$ symmetry of the order
parameter \cite{amin1}. There is no phase winding around these
vortices and therefore no flux quantization is expected. In fact,
the flux trapped in these vortices is much smaller than a flux
quantum. The magnitude of the magnetic field at the maximum
positions is of the order of $10^{-4}$--$10^{-3}G$ in both cases.
This value agrees with the magnitude of the magnetic field
reported in Ref.~\cite{carmi}. In the $d+id^{\prime}$ case, the
magnetic field is strongly peaked at the vortices. The magnetic
field in other spatial points is noticeable only in the vicinity
of the pair-breaking edge (within a few coherence lengths) and is
almost one order of magnitude weaker than the field at the vortex
peaks. This is a result of the superscreening effect which happens
when the spontaneous current is due to the intrinsic angular
moment of the order parameter \cite{amin,amin1}. The total flux
generated by this magnetic field is very small and difficult to
measure. This is in agreement with the magnitude of the flux,
calculated using a different technique, in Ref.~\cite{ZHU}. In the
$d+is$ case, on the other hand, the peak is spread over a wider
region. In fact, in larger systems, the size of this region could
be of the order of the penetration depth. This is because in the
$d+is$ case, the superscreening effect is absent and the only
mechanism to return the current is the Meissner effect, which
happens over the length scale of a penetration depth. The flux
generated by this magnetic field can be large and, therefore,
measurable. This suggests that the flux measured in Ref.~
\cite{carmi} can be generated by a $d+is$ symmetry breaking state
and not by a $d+id^{\prime}$ one.

The spontaneous magnetic moments of mesoscopic islands with
$d+is$ or $d+id^{\prime}$ symmetry can be in principle used as
qubits. Since the ``up" and ``down" states are degenerate
eigenstates of the effective Hamiltonian, tunneling between them
becomes only possible in the presence of external magnetic field,
field gradient, etc., depending on the multipole structure of the
spontaneous current distribution. In order to determine the
tunneling amplitude, a three-dimensional picture should be
considered, which is beyond the scope of this
publication.

\begin{figure}[t]
\epsfysize 5cm \epsfbox[130 190 430 610]{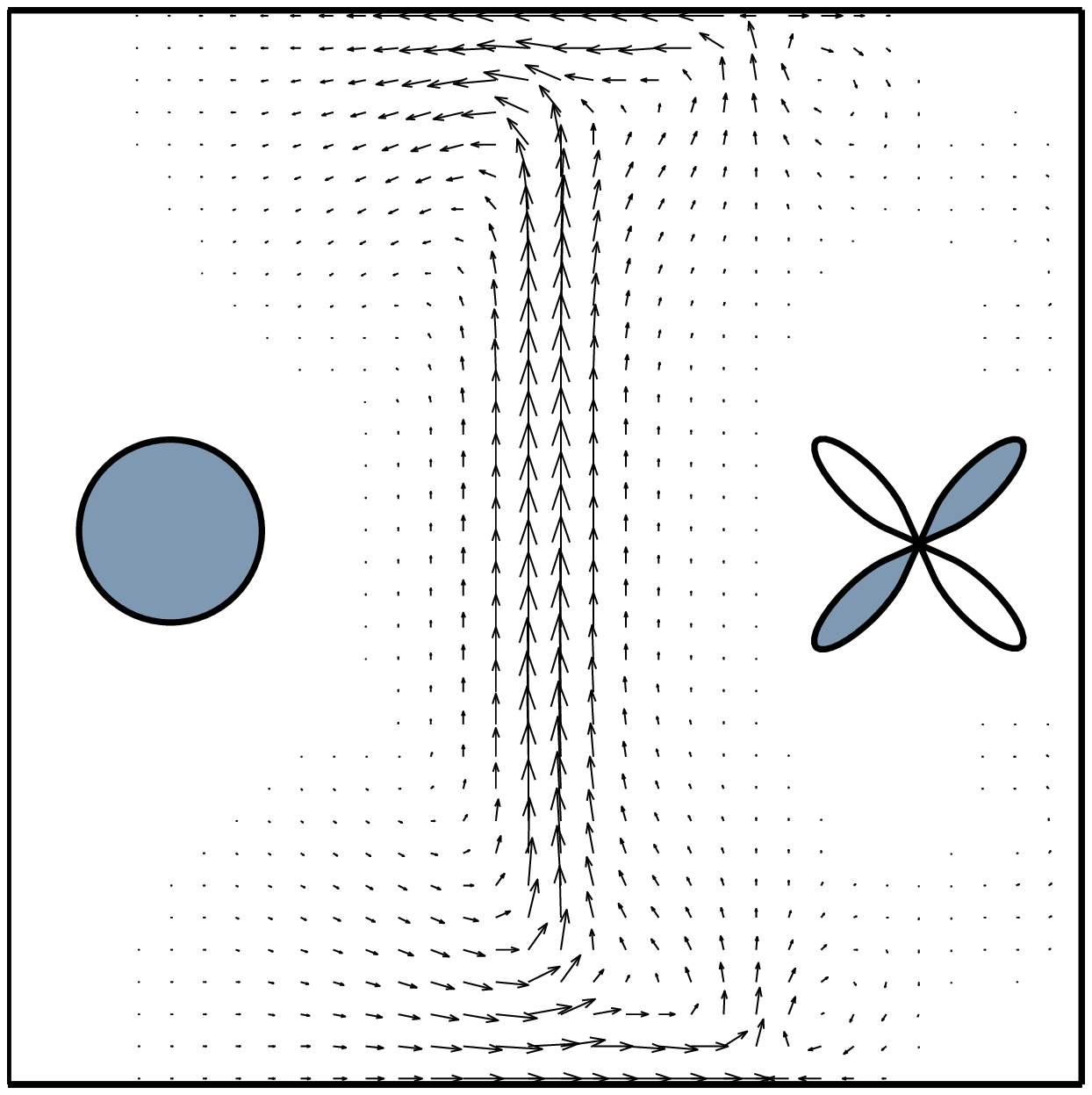} \epsfysize 5cm
\epsfbox[50 200 420 650]{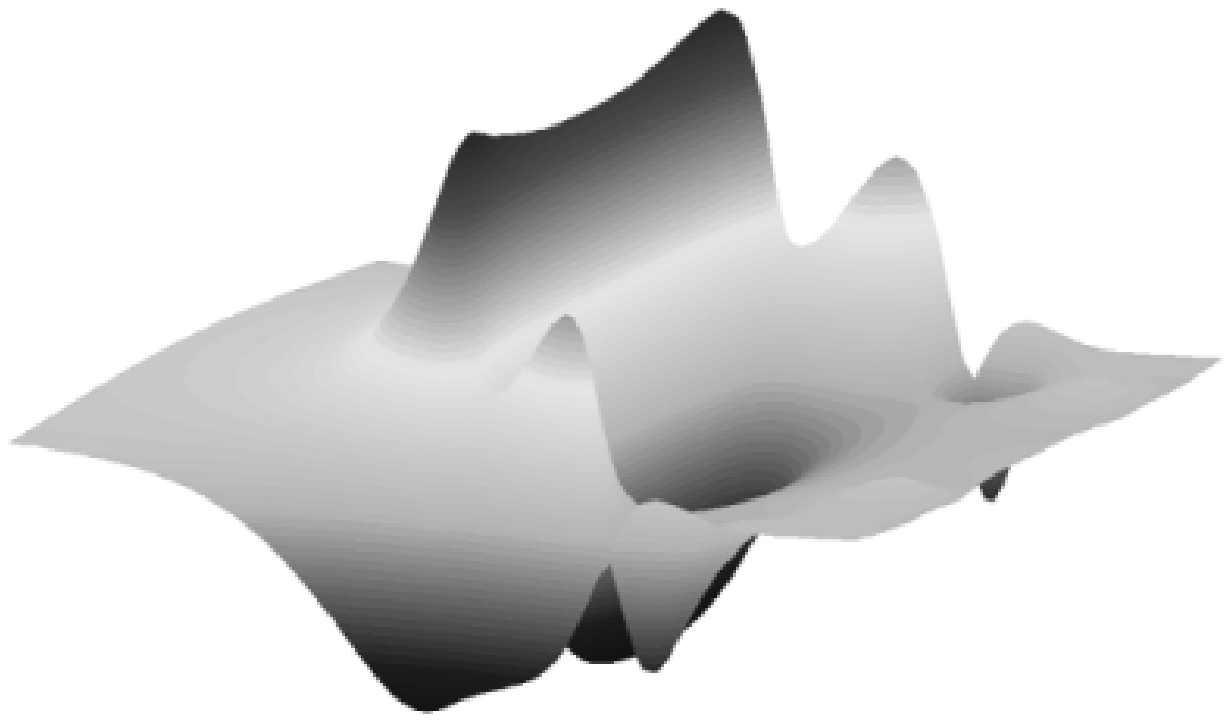} \caption{{\protect\small
Spontaneous current density (left) and magnetic field (right)
distributions for an $s$-$d$ junction. The grain boundary is a
vertical line located in the middle.}} \label{fig3}
\end{figure}

\begin{figure}[t]
\epsfysize 5cm \epsfbox[130 200 410 610]{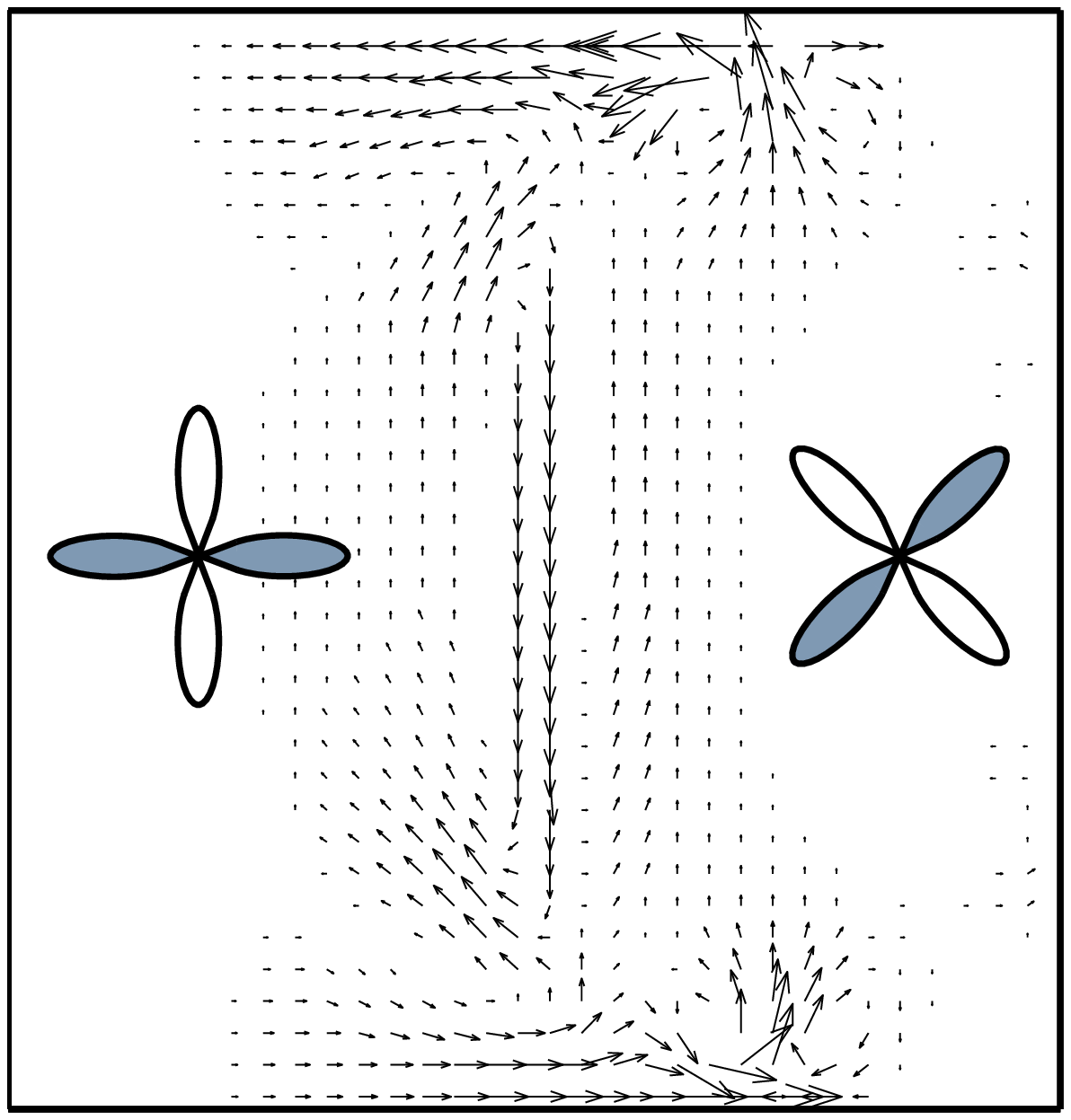} \epsfysize 5cm
\epsfbox[50 200 420 650]{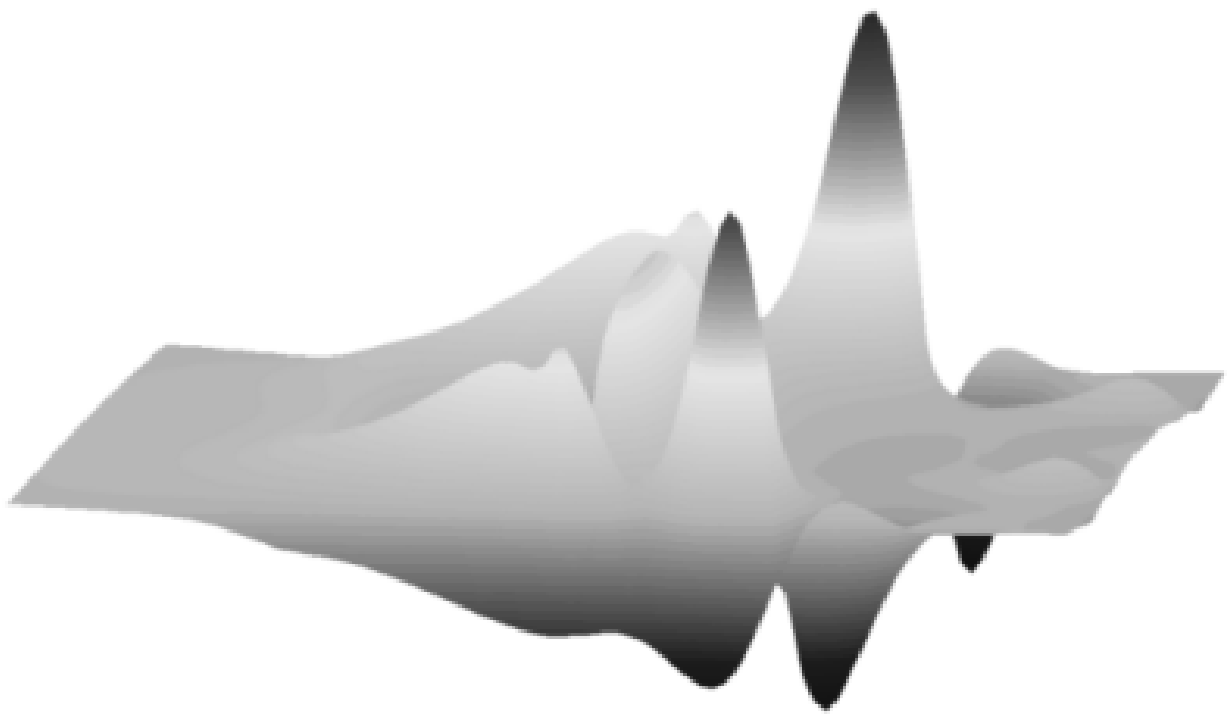} \caption{{\protect\small
Spontaneous current density (left) and magnetic field (right)
distributions for a grain boundary junction between two d-wave
superconductors.}} \label{fig4}
\end{figure}

We also have calculated the spontaneous current and magnetic field
distributions in grain boundary junctions. The geometry of these
systems consists of a square of size $=30\xi_0 \times 30\xi_0$,
divided into two equal parts separated by a grain boundary
junction. The right half of the square is a clean $d$-wave
superconductor with a $45^{\circ }$ orientation of the order
parameter with respect to the boundaries. The left half, on the
other hand, can be either a clean $s$-wave superconductor ($s$-$d$
junction, Fig.~\ref {fig3}), or a clean $d$-wave superconductor
with a $0^{\circ }$-orientation ($d$-$d$ junction,
Fig.~\ref{fig4}). The grain boundary is taken to be perfectly
transparent with no roughness or faceting \cite{hilgenkamp}.
Although this choice of grain boundary does not exactly
correspond to the reality, study of it may provide useful
information for real systems. To be able to compare the two
systems, we assumed that the $s$-wave and $d$-wave
superconductors have the same transition temperatures. We did not
introduce any subdominant order parameter here. However, we
introduced an additional phase difference of $\Delta \phi =\pi
/2$ between the two sides. Such a choice corresponds to the
equilibrium phase difference of the junction at which the total
current passing through the junction is zero \cite{amin1}.
Calculations are done at $T=0.1T_{c}$.

The results of spontaneous current density and magnetic field
distributions are shown in Fig.~\ref{fig3} for the $s$-$d$
junction, and Fig.~\ref{fig4} for the $d$-$d$ junction. Note that
the current distribution is not symmetric with respect to the
grain boundary. For the $s$-$d$ junction of Fig.~\ref{fig3}, the
current has a maximum at the grain boundary and returns through
the bulk of the superconductors on both sides. If the system is
large, the size of the region where the spontaneous current is
non-zero, should be of order of the penetration depth. For the
$d$-$d$ junction (Fig.~\ref{fig4}), on the other hand, the
current changes the direction just within a few coherence length
from the boundary, again due to the superscreening effect
\cite{amin,amin1}. Notice also that near the edges of the system,
on the left side of the grain boundary junction ($0^{\circ
}$-orientation of the order parameter), the current returns along
the diagonal, whereas on the right side ($45^{\circ
}$-orientation) it forms two small vortices and antivortices.
These vortices, which are absent in the $s$-$d$ case, again
reflect the chiral nature of the $d+id^{\prime}$ symmetry;
although the subdominant order parameter is absent in this case,
the correlation functions convey this symmetry near the grain
boundary junction due to the proximity effect \cite{amin1}. The
magnetic field distributions, displayed in Figs.~\ref{fig3} and
\ref{fig4}, is of the order of $10^{-4}G$ at the positions of the
maxima, in both $s$-$d$ and $d$-$d$ junctions. In the case of
$d$-$d$ junction (Fig.~\ref{fig4}), magnetic field is peaked at
the location of vortices. Away from the vortices, the magnetic
field is localized near the grain boundary junction and almost
one order of magnitude smaller. The flux generated by such a
magnetic field is very small. Thus, observation of large flux as
in Ref.~\cite{mannhart} can not be associated with such an
effect. In real systems, due to the faceting effect
\cite{hilgenkamp}, some $\pi$-loops may exist along the grain
boundary junction. They can produce large fluxes, of the same
order as observed in Ref.~\cite{mannhart}.

It is important to emphasize here that the existence of the
spontaneous current at the grain boundary junctions does not
depend on the presence of a subdominant order parameter (unlike
in the previous case, unless if we consider other effects
\cite{Lofwander}). Thus, experiments similar to that of
Ref.~\cite{NV01} (that only probes the mixing of the symmetry of
the order parameter) can not exclude the possibility of the
existence of spontaneous current. Indeed, addition of a
subdominant order parameter will suppress the spontaneous current
at the boundary (see Refs.~ \cite{amin,amin1}).

One should note that in the presence of magnetic field, the
procedure described here is not valid, because a path dependent
phase will be accumulated to $a$ and $b$ functions, and the
relaxation mechanism along the trajectory may no longer hold.
This does not mean that we cannot use our approach to calculate
currents and fields in realistic systems without calculating the
magnetic field due to spontaneous currents self-consistently.
Indeed, it is not difficult to show that the corresponding
Doppler shift is rather small. We saw that for systems with size
$L \sim 30\xi_0 \sim 5 \times 10^{4}$\AA, spontaneous magnetic
field is always smaller than $10^{-3}$G. Therefore, the induced
superfluid momentum is at most $p_s \sim (e/c)HL \sim 10^{-26}
{\rm g}\cdot{\rm cm/ s}$. If we take $\hbar v_F \simeq 1$
eV$\cdot$\AA\cite{FermiVelocity} for the Fermi velocity in
high-T$_c$ cuprates, we obtain $p_s v_F \sim$ 0.1 $\mu$eV $\sim$
1 mK, which is negligible compared to other energy scales in the
problem\cite{aprili}. However, these restrictions may be important
if the system is placed in a strong external magnetic field.

To summarize, we described self-consistent calculations of some
equilibrium properties in finite size superconducting systems. We
calculated distributions of spontaneous current and magnetic
field in different small samples in the presence of mixed order
parameter symmetries. The nature of the mixed symmetry state in
all cases affects the shape of the current and magnetic field
distributions. In particular, the chiral nature of the
$d+id^{\prime}$ states exhibits itself through the appearance of
vortices close to the edges of the system. The vortices are
absent in the $d+is$ cases. The method described here is quite
general and can be applied to any 2D geometry with proper
boundary conditions as long as external magnetic field is not
present. The self-generated magnetic field by the spontaneous
currents, however, is usually very small, i.e., it can be
neglected in most of the practically important cases.

Many discussions with A.~Maassen van den Brink, G.~Rose, and
A.~Yu.~Smirnov are greatly acknowledged.


\begin{references}

\bibitem{sigrist} M. Sigrist and T.M. Rice, J. Phys. Soc.
Jpn. {\bf 61}, 4283 (1992); M. Sigrist, D.B. Bailey, and R.B.
Laughlin, Phys. Rev. Lett. {\bf 74}, 3249 (1995); M. Sigrist,
Prog. Theor. Phys. {\bf 99}, 899 (1998).

\bibitem{huck} A. Huck, A. van Otterlo, and M. Sigrist, Phys. Rev.
B {\bf 56}, 14163 (1997).

\bibitem{barash} Yu.S. Barash, A.V. Galaktionov, and A.D. Zaikin,
Phys. Rev. B {\bf 52}, 665 (1995).

\bibitem{amin}  M.H.S. Amin, A.N. Omelyanchouk, and A.M. Zagoskin, Phys.
Rev. B {\bf 63}, 212502 (2001).

\bibitem{amin1}  M.H.S. Amin, A.N. Omelyanchouk, S.N. Rashkeev, M. Coury,
and A.M. Zagoskin, Physica B {\bf 318}, 162 (2002).

\bibitem{matsumoto}  M. Matsumoto and H. Shiba, J. Phys. Soc. Japan
{\bf 64}, 1703 (1995); {\bf 64}, 3384 (1995); {\bf 64}, 4867
(1995); {\bf 65} 2194 (1996).

\bibitem{Yip} S. Yip, Physical Review B {\bf 52}, 3087 (1995).

\bibitem{Fogelstrom}  M. Fogelstr\"om, D. Rainer, and J.A. Sauls,
Phys. Rev. Lett. {\bf 79}, 281 (1997); M. Fogelstr\"{o}m and S.-K.
Yip, Phys. Rev. B {\bf 57}, R14060 (1998).

\bibitem{Covington}  M. Covington, M. Aprili, E. Paraoanu, L.H. Greene, F.
Xu, J. Zhu, and C.A. Mirkin, Phys. Rev. Lett. {\bf 79}, 277
(1997); L.H. Greene, M. Covington, M. Aprili, E. Badica, and D.E.
Pugel, Physica B {\bf 280}, 159 (2000).

\bibitem{mannhart}  J. Mannhart, {\em et al.} Phys. Rev. Lett. {\bf 77},
2782 (1996).

\bibitem{NV01}  W.K. Neils and D.J. Van Harlingen, Phys. Rev. Lett.
{\bf 88}, 047001 (2002).

\bibitem{carmi}  R. Carmi, E. Polturak, G. Koren, and A. Auerbach, Nature
{\bf 396}, 168 (1998).

\bibitem{tafuri}  F. Tafuri and J.R. Kirtley, Phys. Rev. B {\bf 62}, 13934
(2000).

\bibitem{walter}  H. Walter {\em et al.} Phys. Rev. Lett. {\bf 80}, 3598
(1998).

\bibitem{tzalenchuk}  A.Yu. Tzalenchuk, {\em et al.}, preprint.

\bibitem{pjn} V.B. Geshkenbein, A.I. Larkin, and A. Barone,
Phys. Rev. B {\bf 36}, 235 (1987);  L.N. Bulaevskii, V.V. Kuzii, and
A.A. Sobyanin, JETP Lett. {\bf 25} 290 (1977).

\bibitem{ilichev}  E. I\'lichev, {\em et al.} Phys. Rev. Lett. {\bf 86},
5369 (2001).

\bibitem{schulz} R.R. Schulz, {\em et al.}, Appl. Phys. Lett.
{\bf 76}, 912 (2000).

\bibitem{ACR}  M.H.S. Amin, M. Coury, and G. Rose, IEEE Trans. Appl.
Supercond., in press (cond-mat/0107370).

\bibitem{qubit}  L.B. Ioffe, {\em et al.}, Nature {\bf 398}, 679 (1999);
A. Blais and A.M. Zagoskin, Phys. Rev. A {\bf 61}, 042308 (2000).

\bibitem{volovik} G. E. Volovik,  JETP Lett., {\bf 66},
522 (1997).

\bibitem{book}  V.P. Mineev and K.V. Samokhin, {\em Introduction to
unconventional superconductivity}, Gordon and Breach Science Publishers
(1999).

\bibitem{Eilenberger}  G. Eilenberger, Z. Phys. {\bf 214}, 195 (1968).

\bibitem{Schopohl}  N. Schopohl and K. Maki, Phys. Rev. B {\bf 52}, 490
(1995); N. Schopohl, preprint (cond-mat/9804064).

\bibitem{ZHU}J.-X. Zhu and C.S. Ting, Phys. Rev. B {\bf 60}, R3739
(1999).

\bibitem{hilgenkamp}  H. Hilgenkamp, J. Mannhart, and B. Mayer, Phys. Rev.
B {\bf 53}, 14586 (1996).

\bibitem{Lofwander} T. L\"{o}fwander, V.S. Shumeiko, and G.
Wendin, Phys. Rev. B {\bf 62}, R14653 (2000).

\bibitem{FermiVelocity} A.V. Balatsky and P. Bourges,
Phys. Rev. Lett. {\bf 82}, 5337 (1999) (see note to Ref.[28]).

\bibitem{aprili} M. Aprili, E. Badica, and L.H. Greene, Phys. Rev.
Lett. {\bf 83}, 4630 (1999).

\end{references}
\end{document}